# An Open-Source Framework for Quality-Assured Smartphone-Based Visible Light Iris Recognition

Naveenkumar G. Venkataswamy [a], Yu Liu [b], Soumyabrata Dey [b], Stephanie Schuckers [a], Masudul H. Imtiaz [a]

[a] Department of Electrical and Computer Engineering, Clarkson University, Potsdam, NY, USA

[b] Department of Computer Science, Clarkson University, Potsdam, NY, USA

Corresponding author: E-mail: venkatng@clarkson.edu (Naveenkumar G. Venkataswamy)

## Abstract

Smartphone-based iris recognition in the visible spectrum (VIS) offers a low-cost and accessible biometric alternative but remains a challenge due to lighting variability, pigmentation effects, and the limited adoption of standardized capture protocols. In this work, we present CUVIRIS, a dataset of 752 ISO/IEC 29794-6 compliant iris images from 47 subjects, collected with a custom Android application that enforces real-time framing, sharpness assessment, and quality feedback. We further introduce LightIrisNet, a MobileNetV3-based multi-task segmentation model optimized for on-device deployment. In addition, we adapt IrisFormer, a transformer-based matcher, to the VIS domain. We evaluate OSIRIS and IrisFormer under a standardized protocol and benchmark against published CNN baselines reported in prior work. On CUVIRIS, the open-source OSIRIS system achieves a TAR of 97.9% at FAR = 0.01 (EER = 0.76%), while IrisFormer, trained only on the UBIRIS.v2 dataset, achieves an EER of 0.057\%. To support reproducibility, we release the Android application, LightIrisNet, trained IrisFormer weights, and a subset of the CUVIRIS dataset. These results show that, with standardized acquisition and VIS-adapted lightweight models, accurate iris recognition on commodity smartphones is feasible under controlled conditions, bringing this modality closer to practical deployment.

## 1. Introduction

Iris recognition is a reliable biometric modality due to the permanence and distinctiveness of the iris texture. Patterns form in early childhood and remain stable throughout life, offering stronger invariance than fingerprints (which may wear) or faces (which can change with age and environment) [1]. Iris recognition has been adopted in national-scale identity systems such as India's Aadhaar [2] and Mexico's RENAPO [3], as well as international border control systems.

Most deployed systems rely on near-infrared (NIR) imaging. Melanin is nearly transparent in NIR, which enables reliable capture of fine iris details across both light and dark eyes [1]. NIR illumination also suppresses reflections and reduces ambient interference, yielding consistently high-quality images. Classical pipelines, from Daugman's integro-differential operator to Gabor wavelet encoding [4], were designed for NIR and perform very well under those conditions.

Reliance on NIR hardware creates a deployment bottleneck. Dedicated sensors and active illumination increase cost and complexity in consumer devices. Smartphones include high quality RGB cameras but typically do not support NIR. Prior attempts to add dedicated iris sensors to phones, such as the Galaxy S8/S9, were discontinued due to cost and usability concerns [5]. Consequently, unlike fingerprints or faces [6], [7], iris recognition is largely absent on smartphones.

This work targets smartphone deployment using visible spectrum (VIS) imaging. VIS capture is feasible with existing hardware but remains challenging. Pigmentation reduces contrast in darker irides [8], corneal reflections and glare obscure texture, and handheld capture introduces motion blur and off-axis gaze. Prior VIS datasets often relied on older devices, lacked standardized acquisition procedures, and rarely enforced ISO/IEC 29794-6 quality checks [9]. Reported performance varies widely across studies, and comparisons are often confounded by differences in datasets and protocols. These datasets highlight that data quality and segmentation accuracy are inherently coupled with in VIS iris recognition, motivating our focus on ISO-compliant acquisition and robust lightweight segmentation.

In this work, we present a comprehensive framework for smartphone-based VIS iris recognition. We develop an ISO compliant Android smartphone acquisition app with real-time quality feedback; the first publicly available tool of its kind. We collect the Clarkson University Visible Iris (CUVIRIS) dataset and design a lightweight segmentation model tailored for on-device use. We evaluate CUVIRIS with OSIRIS 4.1 [10] and IrisFormer [11], and we contextualize performance with published CNN results from prior work. In addition, we adapt and train IrisFormer, a transformer-based matcher, for VIS iris recognition. Together, these components provide a systematic and reproducible assessment of VIS iris recognition on smartphones under controlled conditions.

## 2. Background

To study VIS challenges, several datasets have been introduced, spanning from 2005 to 2016. The earliest, UBIRIS.v1 [12] and UBIRIS.v2 [13], were acquired with DSLR cameras. UBIRIS.v1 was collected indoors under controlled conditions. UBIRIS.v2 extended to outdoor sessions and introduced blur, occlusion, and natural lighting. These datasets demonstrated VIS feasibility but were not designed around smartphones.

Later efforts moved to mobile capture. The MICHE-I/II dataset [14] includes images from iPhones, Samsung Galaxy devices, and tablets. VISOB [15] collected more than 4,000 images on three phones, and CSIP [16] examined cross sensor variability with four models. These resources advanced the field, yet common limitations persist older hardware, inconsistent capture protocols, lack of real-time quality checks, and limited consideration of ISO/IEC 29794-6 quality [9]. These gaps motivate our dataset design, which employs modern hardware, enforces ISO compliance, and uses real-time feedback to control acquisition quality. Table 1 summarizes these datasets alongside CUVIRIS.

| Name | Year | Capture device(s) | Environment | Subjects | Images | Custom app | ISO checks |
|---|---|---|---|---|---|---|---|
| CSIP [16] | 2015 | Sony Xperia Arc S, iPhone 4, THL W200, Huawei Ideos X3 | Unconstrained | 50 | 2,004 | No | No |
| MICHE-I/II [14] | 2015 | iPhone 5, Galaxy S4, Galaxy Tab II | Unconstrained | 92 | 3,732 | No | No |
| VISOB [15] | 2016 | iPhone 5S, Galaxy Note 4, Oppo N1 | Indoor/Unconstrained | 550 | ~4,000 | No | No |
| UBIRIS.v1 [12] | 2005 | Nikon E5700 (DSLR) | Indoor (controlled) | 241 | 1,877 | No | No |
| UBIRIS.v2 [13] | 2009 | Canon EOS 5D (DSLR) | Outdoor/Unconstrained | 261 | 11,102 | No | No |
| CUVIRIS (This Work) | 2024 | Samsung Galaxy S21 Ultra | Indoor (ISO-compliant) | 47 | 752 | Yes | Yes |

Table 1: Summary of publicly available VIS iris datasets, compared with the proposed CUVIRIS dataset.

Segmentation remains a persistent challenge in VIS. NIR systems benefited from high iris–sclera contrast and cooperative capture and used circular or elliptical models such as Daugman's integro-differential operator and the circular Hough transform [17]. These methods degrade under VIS conditions where pigmentation, reflections, and occlusions obscure boundaries [18]. Subsequent work proposed probabilistic contour models [18], Fourier-based boundary fitting [19], and landmark-guided refinements [20]. Performance still drops under blur and poor lighting, which motivates lightweight models that are robust in VIS and practical on mobile hardware.

While segmentation defines the quality of iris localization, recognition accuracy ultimately depends on both precise boundary detection and discriminative feature learning. Deep learning improved segmentation quality. Arsalan et al. [21] presented a two-stage CNN for coarse-to-fine segmentation. Osorio-Roig et al. [22] extended to multi-class labeling of iris, pupil, sclera, eyelids, and eyeglasses. Iris R-CNN integrated detection and segmentation in a single pipeline [23]. Multitask frameworks such as IrisParseNet [24] added auxiliary supervision with edge maps and distance transforms, which improved boundary consistency. Many of these models rely on large backbones, for example VGG, ResNet, or DenseNet, which limits their use on mobile devices. More recent work explores lightweight alternatives such as MobileNet-based networks [25], [26] that balance accuracy and efficiency. Our approach follows this direction with a MobileNetV3-based multi-task model optimized for smartphones.

Recognition methods show a similar progression. Classical OSIRIS-style pipelines reported error rates of 7 to 12% on MICHE-I [14], which reflects the difficulty of VIS smartphone capture. Controlled conditions improve performance. For example, commercial SDKs reported high correct match rates [27]. Handcrafted descriptors such as Local Binary Patterns, Binarized Statistical Image Features, and the Weber Local Descriptor yielded 8 to 15% equal error rates on UBIRIS.v2 and MICHE-I, which could be reduced to 8 to 9% with fusion [28]. Later deep methods, including deep sparse filtering [29], patch-based CNNs [30], and DeepIrisNet2 [31], reported 1 to 6% EER. Despite these advances, transformer-based approaches remain under-explored for VIS iris recognition. We address this gap by adapting IrisFormer to the VIS domain and benchmarking it in both closed-set and cross-dataset settings.

Prior work shows steady progress yet persistent challenges. Segmentation quality remains a bottleneck in VIS. Inconsistent capture protocols make comparisons difficult. Acquisition quality is equally important because poor framing, blurry, and uncontrolled lighting can make recognition infeasible. Our study addresses these issues jointly by combining ISO-guided acquisition with real-time quality checks, a lightweight segmentation model for mobile deployment, and a standardized evaluation strategy that includes a transformer-based matcher.

In summary, this work makes the following contributions, all of which are publicly released:

- CUVIRIS dataset: A smartphone VIS dataset collected with a Samsung Galaxy S21 Ultra under an ISO/IEC 29794-6–compliant protocol (752 images from 47 participants). A 10-subject subset is publicly released in line with consent.
- Capture application: An Android app that enforces realtime framing, sharpness assessment, and ISO quality checks for reproducible VIS acquisition. This is the first publicly available quality-assured iris capture tool for smartphones.
- Lightweight segmentation: LightIrisNet, a MobileNetV3-based multi-task model that outputs iris and pupil masks with auxiliary maps (edges, distance transforms, ellipse priors), designed for on-device efficiency.
- Transformer for VIS: Adaptation and training of IrisFormer for VIS recognition with results in closed-set and cross-dataset settings. Our VIS-trained model achieves 0.057% EER on quality-controlled captures.
- Public release: Code, trained models, and the dataset subset, which together form an open-source ecosystem for reproducible VIS iris research on smartphones.

By combining a modern smartphone dataset, a lightweight segmentation model, a standardized evaluation strategy, and open-source resources, this study provides a systematic and reproducible assessment of VIS iris recognition on smartphones.

The remainder of this paper is organized as follows. Section 3 details the acquisition app, dataset, and algorithms used in this study. Section 4 presents experimental results and performance analysis. Section 5 discusses the key findings, limitations, and future work, while Section 6 concludes the paper with a summary of the main contributions.

## 3. Materials and Methods

In this section, we describe the methodology used to develop and evaluate the proposed smartphone-based VIS iris recognition framework. This includes both the dataset construction process and the algorithmic components of the recognition pipeline. We first describe the custom Android application and controlled capture setup used for ISO/IEC-compliant data collection, leading to the construction of the CUVIRIS dataset. We then outline the preprocessing and segmentation strategies designed to ensure reliable iris localization under visible-light conditions. Finally, we present the matching methodology, which includes both a classical hand-crafted pipeline using OSIRIS 4.1 and a transformer-based deep learning matcher, IrisFormer, to enable a fair comparative evaluation under visible-light conditions.

### 3.1. Android Application Development

We developed a mobile software application for the Android platform to enable user-guided capture of high-quality, standards-compliant iris images under visible-light conditions. The app was built in Android Studio using Java, targeting Android API level 21 (Android 5.0 Lollipop) and above to ensure compatibility with most consumer Android smartphones. It functions as a complete data acquisition unit with an integrated system for real-time biometric quality feedback, which is critical to capturing data under consistent conditions for both research and deployment prototyping.

#### 3.1.1. User Interface and Metadata Input

The application features a straightforward user interface (UI) that facilitates the input of participant metadata. Subject ID, Session ID, and Trial Number must be entered before capture. This information is automatically incorporated into each image file name using the convention <subjectID>-<left/right><sessionID>-<trialNumber>.png, ensuring that files remain organized, permanently linked to their metadata, and traceable for reproducibility. The metadata input screen is shown in figure 1(a).

#### 3.1.2. YOLOV3-Tiny Eye and Iris Detection Model

Considering the need to integrate the model into a smartphone while maintaining real-time performance, we adopted the compact object detection architecture YOLOv3-Tiny [32]. This model was chosen for its small parameter count, high speed, and efficiency, featuring 24 layers with 0.557 GFLOPS and capable of processing up to 220 FPS.

Training was carried out using transfer learning with more than 3,000 eye images from the UBIRIS.v1 [33] and UBIRIS.v2 [13] datasets, along with over 750 images from our own smartphone captures. Images were resized to 416×416 pixels and split into 80% training and 20% testing sets. Fine-tuning was performed on a

Nvidia GeForce RTX 3080 GPU, achieving 98.3% training and 97.1% testing accuracy. The resulting weights and configuration were stored in the Hierarchical Data Format (HDF5).

To optimize performance for deployment on mobile devices, we applied post-training quantization, reducing model precision from 32-bit floating point (FP32) to 8-bit integer (int8). This conversion minimized computational load and memory footprint while preserving accuracy. The quantized model was converted from HDF5 to TensorFlow Lite (TFLite) format by mapping floating-point values to discrete integer levels using quantization step size and zero-point parameters.

A comparison between the original and quantized models revealed that the TFLite version reduced file size from 54MB to 33MB and improved inference time on the RTX 3080 GPU from 0.10ms to 0.06ms. On a Samsung Galaxy S21 smartphone equipped with an Adreno 660 GPU, detection time further decreased from 0.26ms to 0.12ms, confirming suitability for real-time deployment on resource-constrained hardware.

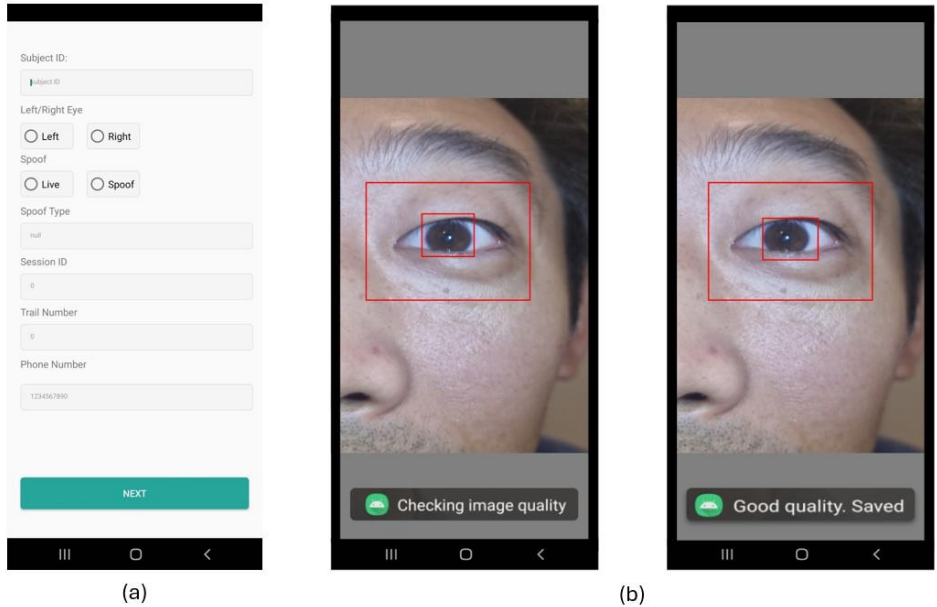

Figure 1: User interface and real-time feedback: (a) metadata input screen for participant details, and (b) live camera preview showing detected eye and iris bounding boxes with real-time quality feedback.

### 3.1.3. Camera Calibration and Frame Scaling

To achieve high-speed inference with the YOLOv3-Tiny detection model, all camera frames were resized to 640×360 pixels from the original sensor resolution. Downstream spatial analysis, such as cropping and quality assessment, required remapping the detected coordinates to the original frame. We therefore introduced a calibration step to recover true landmark positions from resized coordinates.

The scale factors along each axis are defined as:

$$\text{Scale\_factor}_x = \frac{W_{\text{original}}}{W_{\text{resized}}}, \text{Scale\_factor}_y = \frac{H_{\text{original}}}{H_{\text{resized}}}$$

$W_{\text{original}}$, $H_{\text{original}}$ are the original frame dimensions, and $W_{\text{resized}}$, $H_{\text{resized}}$ represent the resized frame (640×360). The computed scale factors from above equation were used to remap detected landmark coordinates $(x,y)$ for spatial consistency.

### 3.1.4. Continuous Autofocus for Optical Sharpness

Consistent eye framing and sharpness are critical for iris recognition. To minimize variability caused by user motion, posture, or distance, the application leverages the smartphone's built-in continuous autofocus mode. Focus is dynamically adjusted on the detected iris region to maintain optimal sharpness during acquisition.

### 3.1.5. Sharpness Estimation

Sharpness is computed in real time using a computationally efficient metric based on the variance of the Laplacian operator. Each frame is converted to grayscale and processed with the Laplacian operator to calculate second-order intensity derivatives. The standard deviation of Laplacian values is used as the sharpness score:

$$S = \sqrt{\frac{1}{N}\sum_{i=1}^{N}(L_i - \mu)^2}$$

where $L_i$ is the Laplacian value for the $i^{th}$ pixel, $\mu$ is the mean Laplacian value, and $N$ is the number of pixels. Images with $S < 70$ were discarded; only frames exceeding this threshold were retained.

### 3.1.6. Compliance with ISO/IEC 29794-6:2015

The ISO/IEC 29794-6:2015 standard defines quantitative metrics for assessing iris image quality. To implement these, we used the opensource BIQT-Iris toolkit [34], which evaluates each frame against the ISO-defined metrics. Thresholds were applied according to the recommended values, summarized in table 2. Only frames meeting all minimum criteria were retained.

| Metric | Range | Recommended value | Applied threshold |
| --- | --- | --- | --- |
| Overall quality | 0–100 | Higher is better | > 70 |
| Grayscale utilization | 0–20 | > 6 | > 6 |
| Iris–pupil concentricity | 0–100 | > 90 | > 90 |
| Iris–pupil contrast | 0–100 | > 30 | > 30 |
| Iris–pupil ratio | 0–100 | 20–70 | > 20 |
| Iris–sclera contrast | 0–100 | > 5 | > 5 |
| Margin adequacy | 0–100 | > 80 | > 80 |
| Pupil-boundary circularity | 0–100 | Higher is better | > 70 |
| Sharpness | 0–100 | Higher is better | > 80 |
| Usable iris area | 0–100 | > 70 | > 70 |

Table 2: Quality metrics defined in ISO/IEC 29794-6:2015 with recommended ranges and applied thresholds in this study.

### 3.1.7. Application Flow

The full on-device acquisition workflow is shown in figure 2. The application operates as a continuous loop with explicit quality checks and user feedback. At the start, the user provides metadata (subject ID, session, and trial), automatically linked to each captured image. The device camera is initialized, and each frame is processed to detect eye and iris landmarks. Coordinates are mapped back to the original resolution, and the eye region is cropped and standardized to 640×480 pixels.

Each frame is assessed for sharpness and ISO/IEC 29794-6 compliance. If the frame meets all thresholds, it is saved with a metadata-based filename; otherwise, the user is prompted to adjust gaze, distance, or eyelid position. This process repeats until eight valid iris images are captured per eye. Figure 1(b) shows the live camera preview showing detected eye and iris bounding boxes with real-time quality feedback.

The entire source code and implementation details are publicly available for reproducibility at
https://github.com/ naveengv7/IrisQualityCapture.

### 3.2 Experimental Setup

All iris images were collected in a controlled indoor environment to ensure repeatable acquisition conditions and minimize illumination variability and reflections. A custom Android application was deployed on a Samsung Galaxy S21 Ultra, using the primary rear wide camera for all captures to maintain consistent optical characteristics. The smartphone was mounted on a height-adjustable tripod in portrait orientation and aligned with each participant's eye level. Participants were seated facing a plain white wall, which provided a uniform background, and the built-in LED flash was enabled to ensure stable illumination. This configuration ensured consistent optical geometry and minimized head movement during capture. The acquisition environment is shown in figure 3.

The application operated in continuous autofocus mode (AF_CONTINUOUS_PICTURE) with a real-time feedback loop that assessed ISO/IEC 29794-6 iris image quality metrics. Frames that failed the quality thresholds were discarded, and participants were prompted to adjust gaze or distance until acceptable images were obtained. On average, the acquisition time for a usable frame was less than 0.1s for light irides and approximately 0.2s for dark irides, likely due to more frequent autofocus adjustments under lower iris-sclera contrast in VIS.

Each eye was captured independently with a target of 8 compliant images per eye, yielding 16 iris samples per subject. All participants provided informed consent and were informed about the study goals, data collection, and retention procedures under an IRB-approved protocol. No personal identifiable information was collected, and anonymization was enforced via the structured filename convention described in Section 3.1.1.

**3.3 CUVIRIS Dataset**

The CUVIRIS dataset was created from 47 volunteer participants aged 18 to 32 years, representing diversity across race, gender, and iris color subgroups. Each subject contributed 8 images per eye, yielding 16 ISO-compliant iris samples per subject. In total, the dataset contains 752 high-quality VIS iris images, all validated against ISO/IEC 29794-6 quality standards.

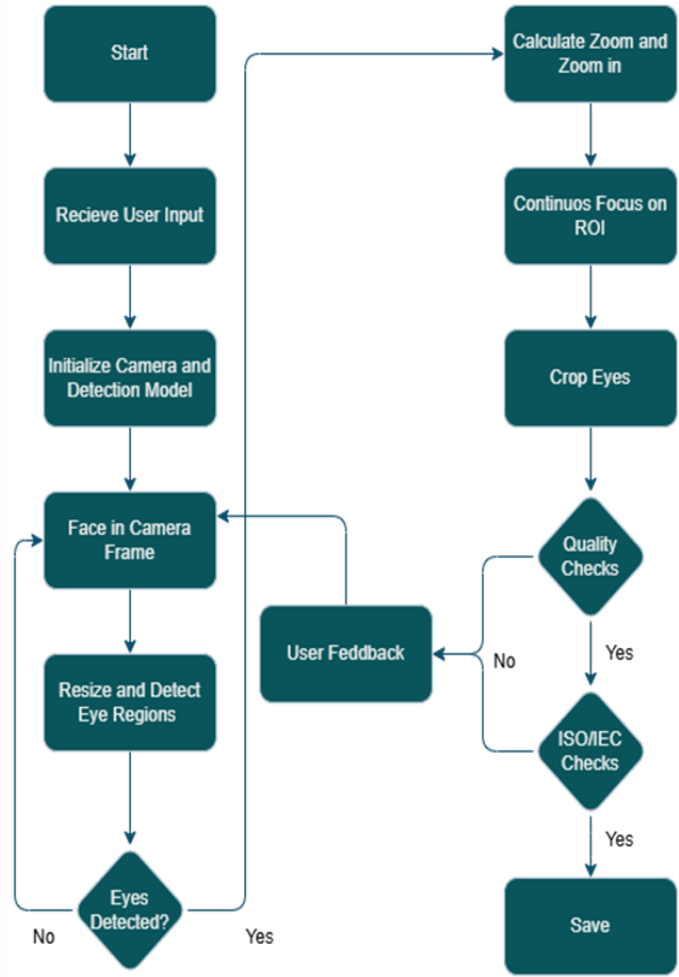

Figure 2: Process flow of the Android application for capturing and processing iris images.

Table 3 summarizes the demographic composition of the CUVIRIS dataset in terms of age, gender, race, and iris color. A balanced representation of light and dark irides enables evaluation under realistic pigmentation variation. Despite the inherent challenges of capturing darker irises under visible illumination, the autofocus and feedback mechanisms of the custom Android application ensured consistent acquisition of high-quality, sharp, and centered iris images.

Representative samples from the dataset are shown in figure 4, illustrating the uniform framing, focus sharpness, and iris visibility achieved across subjects with varying pigmentation levels. Due to participant consent restrictions, the complete 47-subject dataset cannot be publicly shared. However, a 10-subject subset that maintains the same demographic and pigmentation diversity has been made publicly available at https://dx.doi.org/10.21227/4t90-gk02.

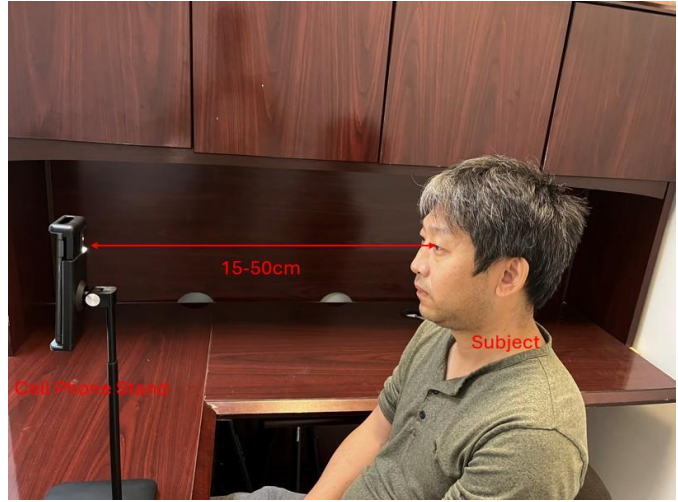

Figure 3: Experimental setup used for data acquisition. The Samsung Galaxy S21 Ultra was mounted on a height-adjustable tripod in portrait orientation at eye level, facing a plain white wall to ensure a uniform background and consistent illumination.

| Category | Count |
| --- | --- |
| Race | |
| Caucasian | 25 |
| Hispanic | 8 |
| Asian | 8 |
| Black | 5 |
| Native American | 1 |
| Gender | |
| Male | 39 |
| Female | 8 |
| Age (years) | |
| 18–25 | 41 |
| 30–32 | 6 |
| Iris color | |
| Light | 21 |
| Dark | 26 |

Table 3: Demographic characteristics of the CUVIRIS dataset in terms of race, age, gender, and iris color.

### 3.4 Segmentation Methodology

Accurate iris segmentation under VIS conditions is challenging due to pigmentation, reflections, and occlusions. Deep multi-task approaches have shown that auxiliary predictions such as edge maps and signed distance transforms (SDTs) improve robustness, but most rely on heavy backbones (e.g., VGG, ResNet, DenseNet) that are unsuitable for mobile deployment. To address this, we propose LightIrisNet, a lightweight segmentation model built on a MobileNetV3 [35] backbone with structured multi-task supervision.

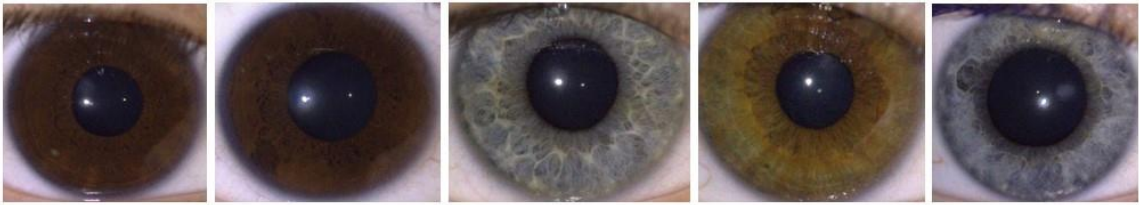

Figure 4: Sample cropped irides images from the CUVIRIS dataset across varying pigmentation levels.

The design balances accuracy with efficiency, targeting real-time VIS iris segmentation on smartphones. An overview of the full pipeline is shown in figure 5: an input eye image is passed through a MobileNetV3-based encoder–decoder with Atrous Spatial Pyramid Pooling (ASPP) [36], producing iris and pupil masks along with auxiliary predictions (boundaries, SDTs, ellipses). These outputs are then used to enforce geometric consistency before the iris texture is normalized for feature extraction and matching.

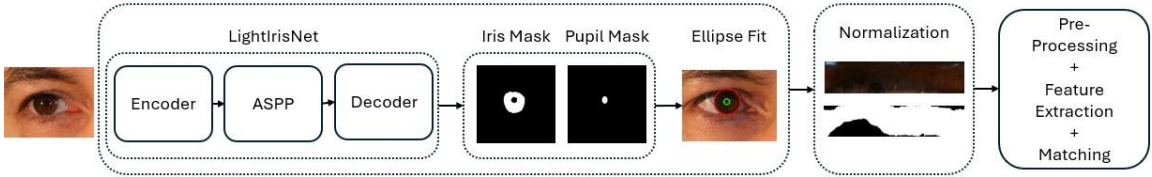

Figure 5: Overview of the proposed pipeline. The MobileNetV3 encoder–decoder with ASPP predicts iris and pupil masks, boundaries, SDTs, and circle parameters. Geometric consistency is enforced before normalization and matching.

### 3.4.1 Model Inputs

Each training sample consists of a color image $I \in \mathbb{R}^{H \times W \times 3}$ paired with multiple annotation-derived targets. From the binary iris and pupil masks, we compute edge maps $E^{iris}$, $E^{pup}$ using Canny detection, which are converted into soft boundary targets $B$ by Gaussian blurring ($\sigma = 1.5$). Signed distance transforms $S^{iris}$, $S^{pup}$ are derived by computing per-pixel distances to the nearest contour, clipped to a fixed radius and normalized into $[-1,1]$. For geometric supervision, ellipses are fitted to binary masks using least squares optimization, yielding parameters

$$\theta = \left(\frac{c_x}{W}, \frac{c_y}{H}, \frac{r_x}{W}, \frac{r_y}{H}, \sin\alpha, \cos\alpha\right)$$

All images are normalized using ImageNet statistics and resized or padded to preserve aspect ratio. This structured representation enables the network to jointly learn pixelwise segmentation and geometric regularization in a unified framework.

**3.4.2. Network Architecture**

The segmentation backbone follows the DeepLabv3+ [37] encoder–decoder paradigm, using MobileNetV3-Large as the encoder. This configuration provides a strong balance between representational capacity and efficiency for mobile deployment. Features are extracted at strides 4, 8, 16, and 32, and global context is aggregated through a depthwise ASPP module with dilation rates of {1, 6, 12, 18}. Decoder fusion proceeds by upsampling ASPP features and combining them with low-level features through 1×1 projections and 3×3 convolutions. The resulting shared feature tensor Z ∈ R$^{H×W×C′}$ feeds lightweight prediction heads. 6 heads produce dense outputs (masks, boundaries, and SDTs), while a global branch regresses ellipse parameters as:

$$\hat{\theta} = \left(\frac{c_x}{W}, \frac{c_y}{H}, \frac{r_x}{W}, \frac{r_y}{H}, \sin\alpha, \cos\alpha\right)$$

where $(c_x, c_y)$ denote the ellipse center, $(r_x, r_y)$ the radii, and $\alpha$ the orientation. The sinusoidal encoding of $\alpha$ ensures numerical stability and avoids discontinuities at $2\pi$.

**3.4.3. Supervision Signals**

Training targets are derived from manual iris and pupil annotations. Binary masks are supervised directly, while contour targets are generated by blurring mask boundaries and weighting them with distance transforms to emphasize near-edge regions. SDTs encode geometric smoothness via signed distances normalized into [−1,1]. Ellipse parameters are estimated via least-squares fitting when feasible and used to train the regression head. This combination of dense and geometric supervision guides the network to respect both local intensity transitions and global anatomical shape regularities.

**3.4.4. Training Objectives**

All outputs are optimized jointly through a weighted sum of task-specific losses:

$$\mathcal{L} = \mathcal{L}_{iris} + \mathcal{L}_{pupil} + \mathcal{L}_{boundary} + \mathcal{L}_{SDT} + \mathcal{L}_{ellipse} + \mathcal{L}_{priors}$$

Here $L_{iris}$ and $L_{pupil}$ denote region losses combining weighted binary cross-entropy with overlap-based terms, with Dice and Tversky loss [38] to mitigate class imbalance:

$$\mathcal{L}_{ii} = \text{BCE}^w(l^{iris}, M^{iris}) + \lambda_{dice}\text{Dice}(l^{iris}, M^{iris})$$

Analogous terms are defined for the pupil, with higher weighting to compensate for its smaller area. $L_{boundary}$ aligns predictions with softened contour targets, $L_{SDT}$ enforces geometric continuity, and $L_{ellipse}$ supervises parametric regression when stable fits are available. The $L_{priors}$ term encodes anatomical consistency through a containment penalty (ensuring the pupil lies within the iris) and an edge–mask alignment constraint. The above equation thus balances pixel-level accuracy, contour sharpness, anatomical plausibility, and geometric compactness.

### 3.4.5. Training Protocol

We aggregate 17,120 VIS iris images from UBIRIS.v1, UBIRIS.v2, MICHE, and CUVIRIS datasets, divided into 13,694 training, 1,713 validation, and 1,713 test samples. These datasets span both smartphone and DSLR devices, ensuring diverse sensor domains. Data augmentations include random scaling, rotations, translations, Gaussian/median blur, and photometric perturbations to mimic real-world artifacts. Mini batches of size 8 are used with dynamic padding to preserve iris geometry. Training employs the AdamW optimizer (learning rate $3 \times 10^{-4}$, weight decay $10^{-4}$) with cosine annealing and mixed precision. Encoder batch-normalization layers are frozen to prevent statistical drift due to cross-dataset variability.

### 3.4.6. Inference and Normalization

During inference, predicted logits are thresholded to obtain binary masks, and ellipses are refitted to the largest connected components. When refitting is unreliable, regressed parameters serve as fallback. Pupil containment is enforced by masking the pupil region within the iris.

Following segmentation, the iris annulus is transformed into a normalized rectangular strip using Daugman's rubbersheet model [17] to standardize scale and rotation across subjects and conditions. This mapping produces normalized iris textures of 512×64 pixels. Occluded regions are masked out to prevent contamination, resulting in ISO-compliant normalized textures ready for subsequent feature extraction and recognition.

### 3.4.7. Ablation Study

To validate design choices, structured ablation experiments were conducted. The baseline MobileNetV3-DeepLabv3+ model predicted only iris and pupil masks. Incorporating lightweight decoder refinements improved boundary precision. Adding SDT and boundary predictions further sharpened edges and stabilized geometry. Replacing the Dice loss with Tversky loss enhanced pupil supervision. Finally, combining all components with limbus specific augmentations yielded the proposed LightIrisNet, which consistently outperformed intermediate configurations.

Trained LightIrisNet models are publicly released for reproducibility at
https://github.com/naveengv7/LightIrisNet.

### 3.5 Iris Preprocessing and Contrast Enhancement

Visible-light iris images often suffer from uneven illumination, specular reflections, and low contrast. These issues are especially pronounced in dark irides, where melanin absorption reduces penetration of visible light and obscures fine patterns. To improve robustness without adding computational overhead, we use a lightweight two-stage preprocessing pipeline applied after segmentation and normalization: (i) red-channel extraction and (ii) gamma correction.

### 3.5.1. Red-channel Extraction

Among the RGB channels, the blue and green bands are most affected by surface reflections and chromatic distortions, while the red channel (620–750 nm) penetrates deeper into the iris stroma. This property makes it less attenuated by melanin and therefore more informative, particularly for dark irides where the other channels often appear flat. Consistent with prior findings in VIS iris imaging [18], we retain only the red channel for subsequent processing.

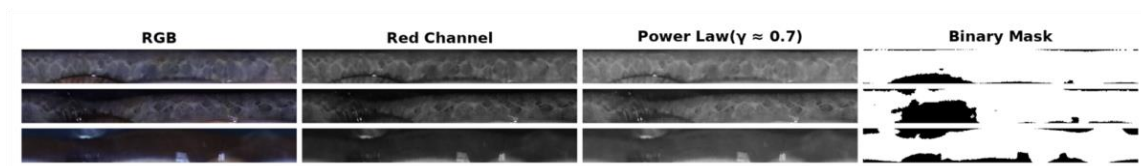

Figure 6: Preprocessing pipeline for normalized iris strips. From left to right: RGB input, red-channel extraction, gamma correction ($\gamma = 0.7$), and the corresponding binary iris mask excluding eyelids, reflections and sclera.

### 3.5.2. Gamma Correction

To further enhance texture visibility, especially in underexposed regions, we apply a power-law gamma transform:

$$I_{\text{out}} = c \cdot I_{\text{in}}^{\gamma}$$

where $I_{\text{in}}$ and $I_{\text{out}}$ are input and output intensities, $c$ is a normalization constant, and $\gamma$ is the exponent. Experiments across datasets with $\gamma$ values between 0.4 and 1.0 showed that very low values amplified noise, while higher ones suppressed useful detail. A setting of $\gamma = 0.7$ provided a consistent tradeoff across both light and dark irides and is adopted uniformly for all datasets to maintain reproducibility and deployment simplicity. Figure 6 shows this preprocessing pipeline.

### 3.6 OSIRIS-based Feature Extraction and Matching

To provide a classical reference point and to validate the fidelity of our captured VIS iris images, we employed OSIRIS 4.1 [10], an open-source implementation of Daugman's iris recognition pipeline [17]. OSIRIS was originally designed for NIR imagery yet applying it directly to VIS-preprocessed normalized iris strips serves two purposes: first, it establishes a baseline for benchmarking; second, if such a traditional NIR optimized system achieves reasonable performance, it confirms that our captured images contain discriminative texture of sufficient biometric quality.

In OSIRIS, iris texture is encoded using a bank of one-dimensional log–Gabor filters, and the resulting phase responses are quantized to form binary iris codes. A binary mask derived from segmentation excludes occluded or noisy regions. During matching, circular bit shifts of up to ±14 positions (about ±15°) are applied to compensate for rotational misalignment, and the lowest fractional Hamming distance across unmasked bits is reported as the similarity score. This representation, together with rotation compensation, has long been considered the gold standard in iris biometrics.

Evaluation followed a standard all-vs-all protocol, with genuine pairs formed from same-eye samples and impostors from cross-subject comparisons. We report results using conventional biometric metrics in Section 4. This baseline establishes that our dataset supports discriminative feature extraction even with classical methods, while also highlighting the limitations of handcrafted coding under VIS-specific degradations such as blur, pigmentation, and uneven illumination.

**3.7 Transformer-based Feature Extraction and Matching**

While OSIRIS provides a classical baseline, its performance in the VIS domain is limited, especially for dark irides and samples affected by blur or reflections. To explore a learning-based alternative, we adopt IrisFormer [11], a recent transformer framework originally developed for NIR iris recognition. The model learns patch-level embeddings that are more resilient to local degradations. Its design includes several features well suited to VIS challenges: relative positional encoding (RoPE) to accommodate rotational offsets after normalization, horizontal pixel–shift augmentation to simulate gaze variability, random token masking to increase robustness against occlusion and glare, and patch-wise sequential matching that preserves fine-grained local texture instead of collapsing everything into a single global descriptor. To our knowledge, this work presents the first systematic evaluation of a transformer-based approach for iris recognition under visible-light conditions, where the model is trained entirely on VIS data (UBIRIS.v2) and tested across multiple smartphone datasets. This adaptation from its original NIR-focused design highlights the potential of transformers to handle VIS-specific difficulties.

We retain the original input representation: each normalized grayscale iris image is divided into non-overlapping patches, linearly projected into embeddings, and processed by a transformer encoder stack. Final descriptors are taken as the sequence of patch embeddings, and matching follows the original patch-wise cosine similarity with an extended distance measure that combines local and global signals.

For reproducibility, we follow the original optimization protocol, initializing from ImageNet-1k weights and training from scratch on UBIRIS.v2, which provides diverse illumination and pigmentation conditions. Training uses AdamW optimization with cosine learning rate scheduling and a margin-based triplet loss, consistent with the original IrisFormer setup. Moderate data augmentation, including horizontal pixel shifts and random patch-level masking, was applied to improve robustness under VIS conditions.

Evaluation mirrors the OSIRIS protocol for fairness. Models trained on UBIRIS.v2 are tested on UBIRIS.v1, MICHE, and CUVIRIS in exhaustive all-vs-all verification (same-eye genuine pairs; cross-subject impostors). Results are reported using standard biometric error rates (TAR, FAR, EER), with additional score-distribution statistics presented in section 4. Trained VIS-transformer models are released for reproducibility at https://github.com/naveengv7/Vis-IrisFormer.

# 4. Results

This section presents the quantitative and qualitative results obtained using the proposed LightIrisNet for segmentation, along with recognition results based on OSIRIS and IrisFormer.

## 4.1. LightIrisNet Segmentation Performance

We evaluate the proposed MobileNetV3-based LightIrisNet segmentation framework on a diverse collection of iris datasets, including UBIRIS.v1, UBIRIS.v2, MICHE, and CUVIRIS. Results are reported in terms of Intersection over Union (IoU), Dice coefficient, and boundary error ($E_1$). IoU and Dice are computed per mask and averaged across all test images. The boundary error is computed as the mean absolute difference between the predicted probability map and the ground-truth binary mask, averaged over all pixels, corresponding to an $L_1$ error bounded in [0,1].

On the combined evaluation across all four datasets, LightIrisNet achieves an IoU of 0.905 and Dice of 0.9386 for the iris, with boundary error $E_1$ = 0.01. For the pupil, we obtain IoU 0.8951, Dice 0.9184, and $E_1$ = 0.002. These results indicate that the proposed model produces masks that closely match ground-truth annotations, with particularly low boundary errors that are important for reliable normalization.

To further assess generalization, the model is evaluated separately on each dataset. On UBIRIS.v2, the model achieves iris Dice 0.941 and pupil Dice 0.928. On UBIRIS.v1, we observe Dice 0.936 (iris) and 0.912 (pupil). On MICHE, which includes images from multiple mobile devices under varied conditions, the Dice reaches 0.923 (iris) and 0.917 (pupil). The highest performance is obtained on CUVIRIS, our controlled VIS dataset, with iris Dice 0.954 and pupil Dice 0.937.

Table 4 compares our results with representative segmentation methods from literature. Learning-based models such as IrisParseNet [24] and Deep Multi-Task Attention Networks [22] improved segmentation accuracy by incorporating multi-task supervision and attention mechanisms but relied on heavy backbones such as VGG-16 and ResNet-50. IrisDenseNet [39] further increased the iris Dice to approximately 0.972 using DenseNet connections, although pupil segmentation was not addressed. In contrast, our lightweight LightIrisNet achieves competitive Dice scores (0.941 for iris and 0.928 for pupil) while being evaluated across multiple datasets simultaneously.

Figure 7 shows representative segmentation outputs on MICHE, UBIRIS.v2, UBIRIS.v1, and CUVIRIS. Across challenging conditions such as specular reflections, eyelid and eyelash occlusions, eyeglasses, off-axis gaze, and low iris–sclera contrast; the predicted masks remain well-localized, and the fitted circles align closely with ground-truth boundaries. The resulting 512×64 normalized strips and binary masks confirm consistent unwrapping with occlusions effectively suppressed, demonstrating the robustness of the proposed multi-task supervision under diverse visible-light acquisition conditions.

| Dataset | Method | Metric | Reported result |
| --- | --- | --- | --- |

| Dataset | Method | Metric | Value |
|---|---|---|---|
| UBIRIS.v1 | LightIrisNet | IoU / Dice (iris) | 0.925 / 0.936 |
|  | LightIrisNet | IoU / Dice (pupil) | 0.892 / 0.912 |
| UBIRIS.v2 | Wang et al. (IrisParseNet) [24] | AUC@0.3 | 0.2865 |
|  | Arsalan et al. (IrisDenseNet) [39] | IoU / Dice (iris) | 0.946 / 0.972 |
|  | LightIrisNet | IoU / Dice (iris) | 0.923 / 0.941 |
|  | LightIrisNet | IoU / Dice (pupil) | 0.905 / 0.928 |
| MICHE | Wang et al. (IrisParseNet) [24] | AUC@0.3 | 0.2903 |
|  | Arsalan et al. (IrisDenseNet) [39] | IoU / Dice (iris) | 0.933 / 0.965 |
|  | LightIrisNet | IoU / Dice (iris) | 0.907 / 0.923 |
|  | LightIrisNet | IoU / Dice (pupil) | 0.898 / 0.917 |
| CUVIRIS | LightIrisNet | IoU / Dice (iris) | 0.937 / 0.954 |
|  | LightIrisNet | IoU / Dice (pupil) | 0.906 / 0.937 |

Table 4: Benchmarking iris segmentation performance on UBIRIS.v1, UBIRIS.v2, MICHE, and CUVIRIS. Only segmentation results are reported. Metrics are shown as reported in the original papers.

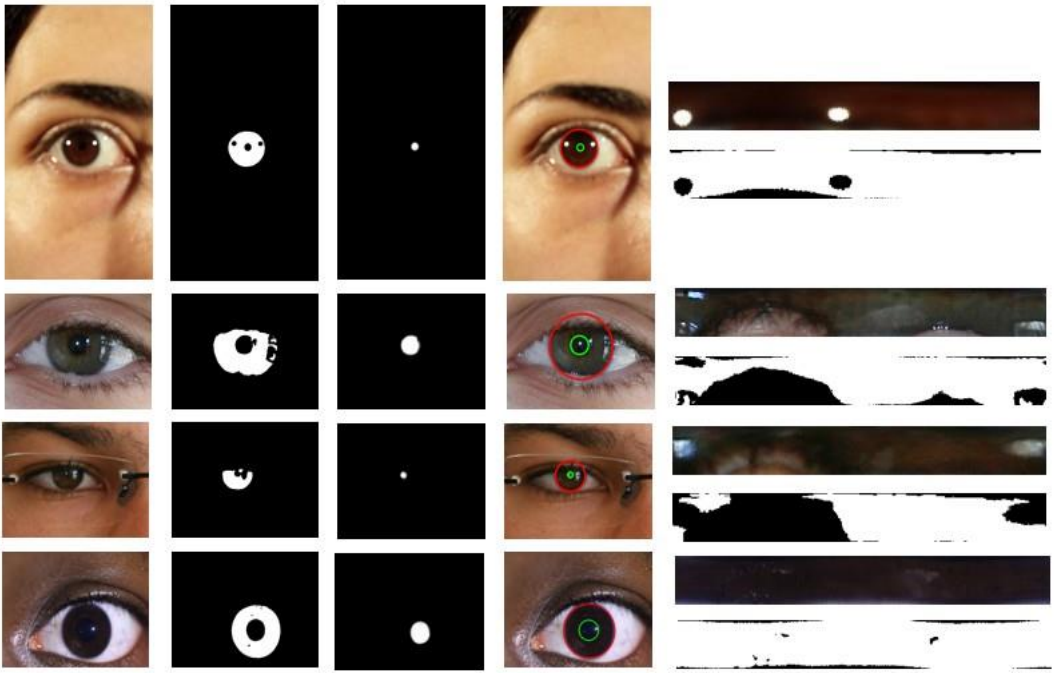

Figure 7: Qualitative segmentation outputs across heterogeneous VIS datasets. Columns show original image, predicted iris and pupil masks, circle fits, and normalized iris strip with binary mask. Rows correspond to MICHE, UBIRIS.v2, UBIRIS.v1, and CUVIRIS.

### 4.2. OSIRIS Results

To establish a classical performance baseline, the OSIRIS 4.1 open-source iris recognition system was evaluated on the CUVIRIS dataset under three experimental conditions. First, the All Subjects set includes all 47 participants and reflects overall performance. To study the effect of pigmentation in visible light, the dataset was further divided into two groups: (i) a Dark-Eyed Subset of 26 participants with brown or black irides and (ii) a Light-Eyed Subset of 21 participants with blue, green, or hazel irides. This division is motivated by the known impact of melanin on iris texture visibility and segmentation reliability under VIS illumination.

On the All Subjects set, OSIRIS achieved an EER of 0.76% at threshold 0.590 and a TAR of 97.9% at FAR = 0.01%. Genuine and impostor distributions were clearly separated, confirming that the CUVIRIS captures provide a discriminative texture of sufficient quality even for a system designed for NIR imagery. Performance declined on the Dark-Eyed Subset, where the EER rose to 1.29% and TAR at FAR = 0.01% dropped to 97.1%. In contrast, the Light-Eyed Subset achieved perfect separation with EER = 0.00% and TAR = 100%. Figure 9 illustrates this effect: the dark-eyed curve shifts upward relative to the all-subject curve, while the light-eyed subset is omitted because no errors were observed.

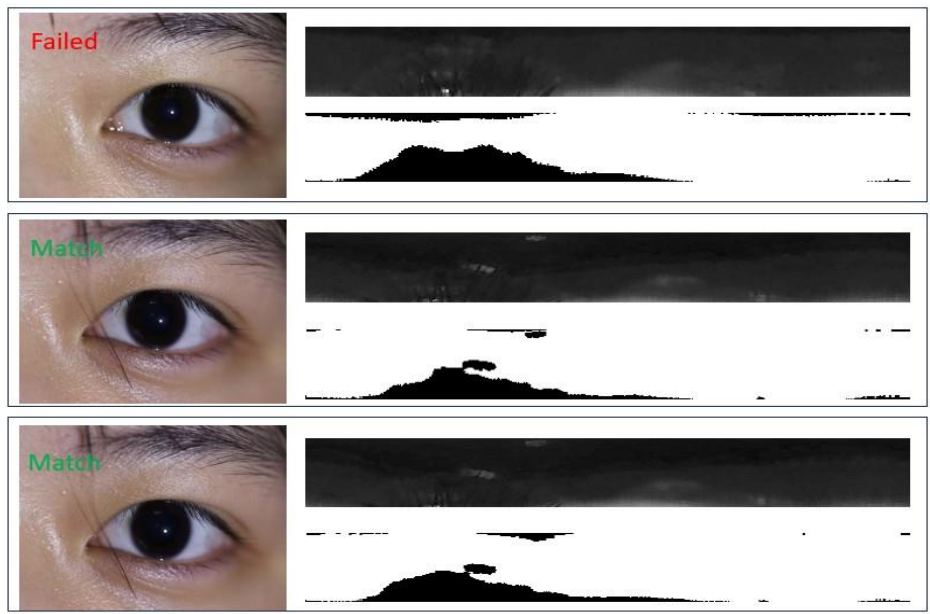

Figure 8: Representative OSIRIS failure case on CUVIRIS (Subject 016, left eye). Top: failed sample showing blurred iris boundaries and uneven sharpness. Bottom: correctly matched same-eye samples with uniform focus and texture.

A small number of failures were observed among dark-eyed subjects. Figure 8 illustrates a representative case (Subject 016, left eye): one sample failed to match its same-eye mates due to motion blur and uneven focus across the normalized strip, compounded by minor illumination non-uniformity. Qualitative inspection showed blurred inner and outer boundaries with band-limited sharpness, leading to unstable phase bits and reduced matching reliability.

Qualitative inspection of errors showed that false rejects were mostly caused by localized degradations such as motion blur, uneven sharpness in the normalized strip, and minor illumination non-uniformities. These issues destabilized the phase encoding, leading to unreliable iris codes. Such effects were noticeably more common in dark irides, where reduced reflectance amplifies the impact of blur and poor focus. Taken together, these findings confirm that smartphone-based VIS iris images of sufficient quality can support classical recognition, while also underscoring pigmentation as the main challenge that motivates learning-based methods.

### 4.3. IrisFormer Results

To assess the transferability of transformer-based recognition into the visible-light (VIS) domain, we retrained IrisFormer on UBIRIS.v2 and evaluated it across UBIRIS.v1, UBIRIS.v2, MICHE, and CUVIRIS. This setup enables direct benchmarking against both handcrafted and CNN-based VIS recognition methods, while highlighting the impact of controlled versus unconstrained acquisition conditions.

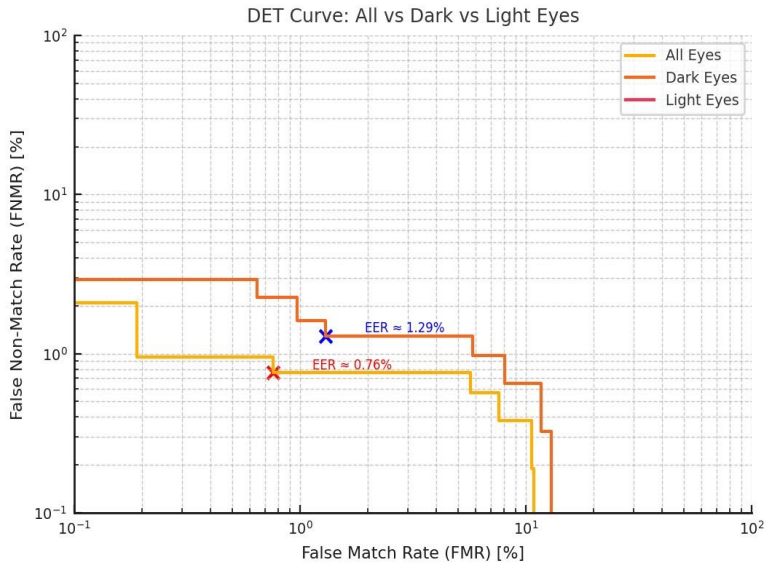

Figure 9: DET curves for OSIRIS on CUVIRIS: All Subjects and Dark-Eyed subsets. The Light-Eyed subset achieved perfect separation and is therefore not plotted.

On UBIRIS.v2, IrisFormer achieved an EER of 5.12%. While this dataset includes frequent blur, glare, and off-angle captures, the model produced stable embeddings and was competitive with CNN-based approaches such as SCNN [30] (5.62%) and DeepIrisNet2 [31] (8.51%). Compared to handcrafted descriptors like DSF [29] (6.1%) or Raghavendra & Busch [28] (∼8-15%), IrisFormer provided clear improvements, underscoring the advantage of learned patch embeddings in challenging VIS conditions.

| Dataset | Method | EER (%) |
|---|---|---|
| UBIRIS.v1 | Iris Former | 1.15 |
| UBIRIS.v2 | Handcrafted: Raghavendra & Busch [28] | ~8.15 |
| | Handcrafted: Raja et al. (DSF) [29] | 6.10 |
| | CNN: Zhao & Kumar (SCNN) [30] | 5.62 |
| | CNN: Gangwar et al. (DeepIrisNet2) [31] | 8.51 |
| | IrisFormer (closed-set) | 5.12 |
| MICHE -I | Handcrafted: De Marsico et al. (OSIRIS baseline) [14] | ~7-12 |
| | Handcrafted: Raghavendra & Busch [28] | ~9-14 |
| | CNN: Raja et al. (DSF-CNN variant) [29] | 5.70 |
| | CNN: Zhao & Kumar (SCNN) [30] | ~5-7 |
| | IrisFormer (cross-dataset) | 8.78 |
| CUVIRIS | OSIRIS (All Subjects) | 0.76 |
| | IrisFormer (Cross-dataset) | 0.057 |

Table 5: Recognition results (EER %) across UBIRIS.v1, UBIRIS.v2, MICHE-I, and CUVIRIS. Methods are grouped by type. Protocol differences follow the original papers.

On UBIRIS.v1, which exhibits fewer degradations than UBIRIS.v2, IrisFormer obtained an EER of 4.15%. Genuine and impostor distributions were better separated than in UBIRIS.v2, showing that transformer-based embeddings generalize across sensors when the acquisition quality improves.

The MICHE dataset posed the greatest challenge, as it combines three heterogeneous mobile devices under uncontrolled illumination. Here, IrisFormer reached an EER of 8.78%. While weaker than its performance on DSLR-based datasets, it still matched the range of handcrafted baselines [14], [28] and remained competitive with CNN-based systems, which report EERs between 5-7%.

Table 5 provides the dataset-wise comparison across VIS benchmarks. On CUVIRIS, collected under controlled conditions with autofocus and subject guidance, IrisFormer achieved near perfect separation with an EER of just 0.057%. Figure 10 shows the DET curves. We further report distributional and operating-point statistics (GMean, GSTD, IMean, ISTD, $d'$, AUC, ZeroFMR, ZeroFNMR). As shown in table 6, CUVIRIS exhibits the strongest separation (AUC = 1.0000, $d'$ = 4.808), with highly consistent genuine scores (GMean = 0.9312, GSTD = 0.0255). UBIRIS.v1 (AUC = 0.9920, $d'$ = 3.195) and UBIRIS.v2 (AUC = 0.9898, $d'$ = 3.002) follow closely, while MICHE shows the weakest separation (AUC = 0.9717, $d'$ = 2.386), consistent with its higher EER. This surpasses both CNN-based and handcrafted baselines, highlighting that when acquisition quality is assured, transformer-based embeddings can deliver state-of-the-art recognition in VIS. Notably, samples that challenged traditional pipelines (e.g., blur or uneven illumination) were successfully matched by IrisFormer, demonstrating resilience to VIS-specific degradations.

| Dataset | GMean | GSTD | IMean | ISTD | $d'$ | AUC | ZeroFMR (%) | ZeroFNMR (%) |
|---|---|---|---|---|---|---|---|---|
| MICHE | 0.7671 | 0.0575 | 0.6382 | 0.0503 | 2.386 | 0.9717 | 73.98 | 87.29 |
| UBIRIS.v1 | 0.8772 | 0.0628 | 0.6692 | 0.0673 | 3.195 | 0.9920 | 30.68 | 75.88 |
| UBIRIS.v2 | 0.7497 | 0.0381 | 0.5928 | 0.0634 | 3.002 | 0.9898 | 61.91 | 61.82 |
| CUVIRIS | 0.9312 | 0.0255 | 0.6479 | 0.0794 | 4.808 | 1.0000 | 0.26 | 0.09 |

Table 6: Comprehensive performance summary of IrisFormer across VIS iris datasets. Higher $d'$ and AUC indicate stronger separability.

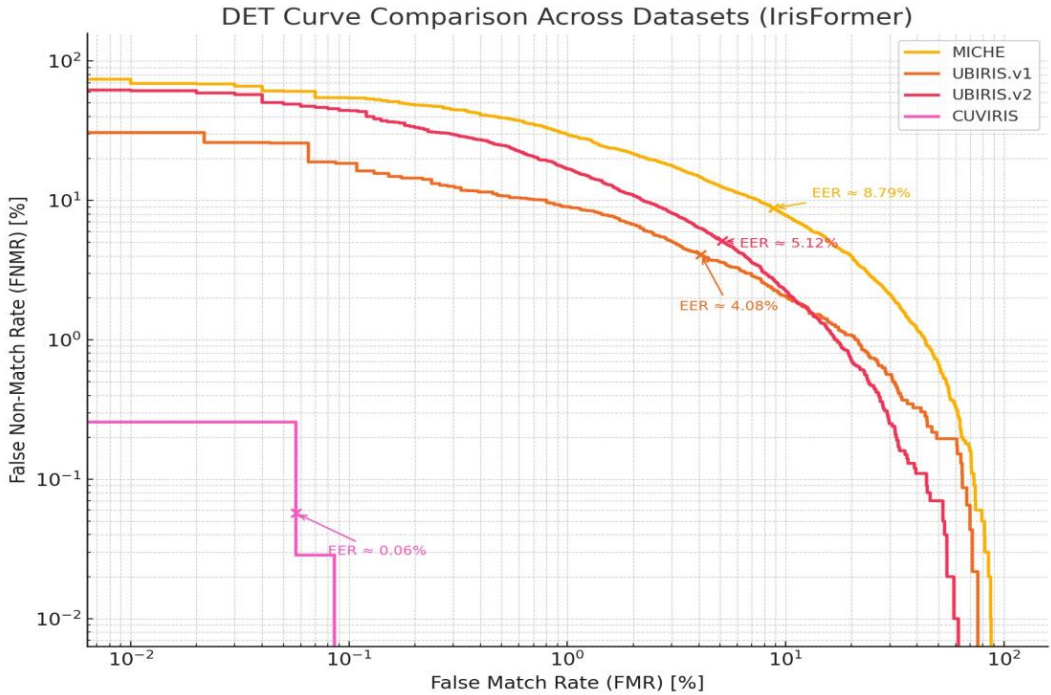

Figure 10: DET curves for IrisFormer across UBIRIS.v1, UBIRIS.v2, MICHE, and CUVIRIS datasets. EER operating points are annotated.

## 5. Discussion

This study demonstrates that visible-light iris recognition on commodity smartphones is technically feasible when acquisition is standardized and quality is actively enforced. Previous VIS datasets such as UBIRIS.v1/v2 and MICHE established important baselines, but they lacked consistency in capture protocols and rarely reflected mobile device conditions. By introducing the CUVIRIS dataset, collected with real-time ISO/IEC 29794-6 checks, and by evaluating both a handcrafted NIR-oriented baseline (OSIRIS) and learning-based approaches adapted to VIS, we provide an integrated benchmark for assessing the modality under smartphone conditions.

The results highlight several scientific insights. First, the long-recognized difficulty of recognizing dark irides in VIS can be substantially mitigated when image quality is controlled at capture. OSIRIS, originally designed for NIR imagery, reached a TAR of 97.9% at FAR = 0.01 even on CUVIRIS, with only modest degradation between light- and dark-eyed subsets. This suggests that smartphone sensors, when coupled with active quality enforcement and basic contrast enhancement, can recover discriminative texture information across pigmentation levels. Second, transformer-based embeddings proved to generalize more effectively than handcrafted codes or CNNs. IrisFormer, trained only on UBIRIS.v2, transferred successfully to UBIRIS.v1, MICHE, and CUVIRIS, achieving a near-perfect EER of 0.057% on the latter. These findings indicate that attention-based models are well-suited to VIS specific degradations such as blur, reflections, and pigmentation differences, and that training on one VIS dataset can provide transferable representations for others.

From a practical perspective, the Android application and quality-controlled acquisition pipeline ensured that only ISO-compliant frames were retained, providing consistent input across subjects and sessions. This demonstrates a path toward reproducible smartphone-based VIS iris research without overstating claims of being the "first" benchmark. The lightweight LightIrisNet segmenter further reinforces the feasibility of on-device deployment: while not achieving the very highest Dice scores reported by heavier CNNs, it balances accuracy, efficiency, and cross-dataset robustness in a form factor suitable for mobile execution.

At the same time, several limitations should be acknowledged. The CUVIRIS dataset is modest in scale, comprising 47 participants, and is demographically imbalanced (39 male, 8 female). All samples were acquired indoors using a single flagship smartphone (Galaxy S21 Ultra rear camera) under controlled conditions and thus do not reflect the variability introduced by outdoor illumination, front-facing cameras, or mid-range devices. Furthermore, this work focused on cooperative verification scenarios, leaving open questions about open-set identification, presentation attack detection, and comprehensive profiling of latency and energy use on-device.

Future work should therefore expand along multiple dimensions. Enlarging CUVIRIS to include more participants with greater demographic diversity, multiple smartphones of varying tiers, and data collected under uncontrolled and outdoor conditions would strengthen its role as a benchmark. Algorithmically, lightweight transformer variants, distillation techniques, and quantization strategies are promising avenues for enabling real-time matching directly on smartphones. Finally, integrating acquisition, segmentation, enhancement, and recognition into a unified mobile application would serve as a compelling demonstration of end-to-end feasibility.

## 6. Conclusion

This study shows that visible-light iris recognition on smartphones is technically feasible when capture is standardized and image quality is actively enforced. Through the CUVIRIS dataset, a dedicated Android acquisition app, and a lightweight segmentation model, we provide a reproducible framework for evaluating this modality under mobile conditions. Our experiments demonstrate that transformer-based embeddings achieve strong recognition accuracy, outperforming the classical OSIRIS pipeline and showing greater robustness than prior CNN-based approaches, particularly for dark-eyed subjects. At the same time, the dataset remains limited in size and demographic balance, and cross-dataset evaluations highlight domain shifts as a persistent challenge. Expanding data diversity and advancing efficient on-device models will be key to making smartphone-based iris recognition practical in real-world use.

## 7. Data Availability

All datasets, trained models, and source code associated with this work are available via the unified repository VISIrisHub at https://github.com/naveengv7/VISIrisHub.

## 8. Conflicts of Interest

The authors declare no competing interests.

## 9. Funding Statement

This work was supported in part by the Center for Identification Technology Research (CITeR) and the National Science Foundation under Grant Nos. 1650503, 2501916, and 2413228.

## 10. Acknowledgements

The authors thank all participants for supporting this research.